\documentclass{PoS}

\def\pythia{{\sc Pythia}}
\def\cascade{{\sc Cascade}}
\def\herwig{{\sc Herwig}}

\title{
\vspace*{-2cm}
\begin{flushright}
{\rm \small
OUTP-08-04-P}\\
\end{flushright}
\vspace*{2cm}
Three-jet DIS final states from\\ k$_\perp$-dependent parton showers}

\ShortTitle{Three-jet final states}

\author{{F.~Hautmann}\\
        University of Oxford\\
        E-mail: \email{hautmann@thphys.ox.ac.uk}}
	
\author{H.~Jung\\
        DESY\\
        E-mail: \email{jung@mail.desy.de}}

\abstract{Experimental measurements of angular correlations in  
three-jet final states have recently been performed in DIS.  
  Next-to-leading-order QCD results for these observables are 
 affected by large  theoretical uncertainties in the 
kinematic region of the data. 
We discuss  the  effects of 
multiple  QCD radiation at higher order 
using parton-shower methods based on 
transverse-momentum dependent parton distributions 
and matrix elements. We present Monte-Carlo results for 
azimuthal two-jet and three-jet distributions, and discuss 
 the comparison with 
experimental data. }  

\FullConference{8th International Symposium on Radiative Corrections \\
        October 1-5, 2007\\
        Florence, Italy}

\begin{document}

The study of hadronic final states with multiple jets at the 
Tevatron, HERA and LHC colliders relies  both on perturbative 
 multi-jet calculations  
  (see~\cite{bernhouches} for a recent overview) and 
 on  realistic  event simulation  by 
 parton-shower Monte-Carlo generators   (see  
e.g.~\cite{mlmhoche,heralhcproc}).
These complex processes, 
characterized by multiple hard scales, 
are  potentially   sensitive 
to effects of QCD  initial-state  radiation 
 that depend on the finite transverse-momentum 
tail of partonic matrix elements and 
distributions~\cite{heralhcproc,hj_rec}. 
In perturbative multi-jet 
calculations 
truncated to fixed order in $\alpha_s$~\cite{bernhouches},   
 finite-k$_\perp$ contributions 
 are taken into account  partially, 
order-by-order,  through higher-loop  corrections. 
On the other hand, 
in standard shower Monte-Carlos  such as  
\herwig~\cite{herwref} and 
\pythia~\cite{pythref},  
based on collinear evolution of the initial-state jet, 
finite-k$_\perp$ contributions are not included, but rather  
  correspond to corrections~\cite{skewang,hef} to the 
angular or transverse-momentum ordering  implemented in the 
parton-branching   algorithms. 

In this article we discuss the role of 
 initial-state  radiative effects in the study of 
 angular  correlations 
for  multi-jet processes, concentrating on the case 
of three-jet DIS production, for which new experimental 
data have 
recently appeared~\cite{zeus1931} and 
next-to-leading-order (NLO) calculations 
are available~\cite{nagy}. 
In the first part  we recall the experimental 
 results~\cite{zeus1931}  and observe 
 that while 
inclusive jet cross sections are reliably predicted 
by NLO perturbation theory, jet correlations are affected by 
significant higher-order corrections, 
increasing 
as the momentum fraction $x$ decreases, 
 giving large 
theoretical uncertainties  at NLO.  
 Sizeable   contributions  
arise from regions~\cite{hj_rec} with 
 three well-separated hard jets 
in which the partonic  lines along the decay chain 
in the initial state  are not ordered in transverse momentum.  
In the second part we move on to the description of 
multi-jet correlations by parton showers 
that incorporate finite-k$_\perp$ corrections to
the   transverse-momentum ordering~\cite{hj_ang}.   The results 
compare well with experimental data, and give quite 
  distinctive features of the associated jet distributions 
compared 
 to  standard showers such as  \herwig.

\section{Jet distributions from NLO calculations}
\label{sec1}

\begin{figure}[htb]
\vspace{50mm}
\includegraphics{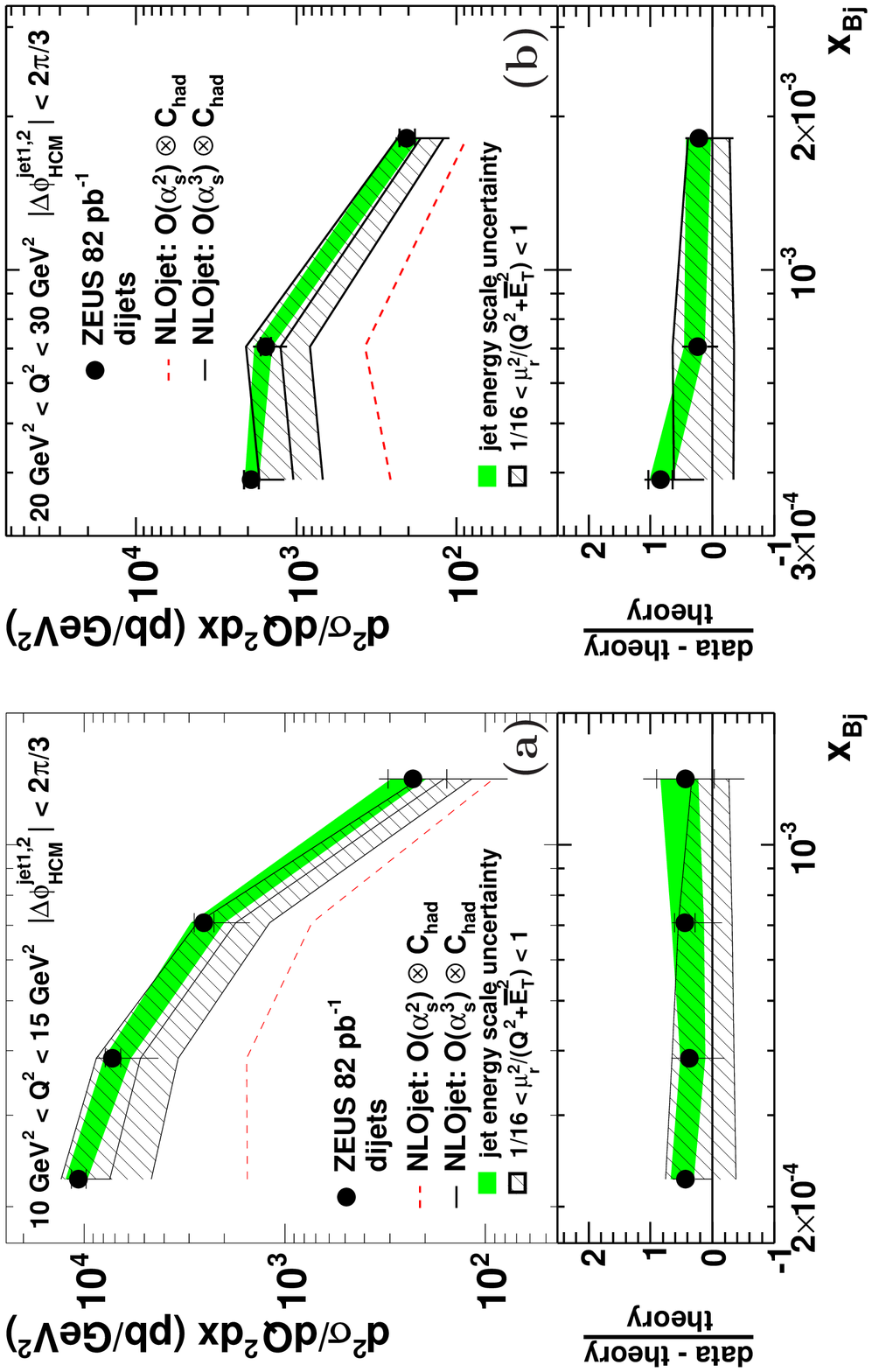}
\includegraphics{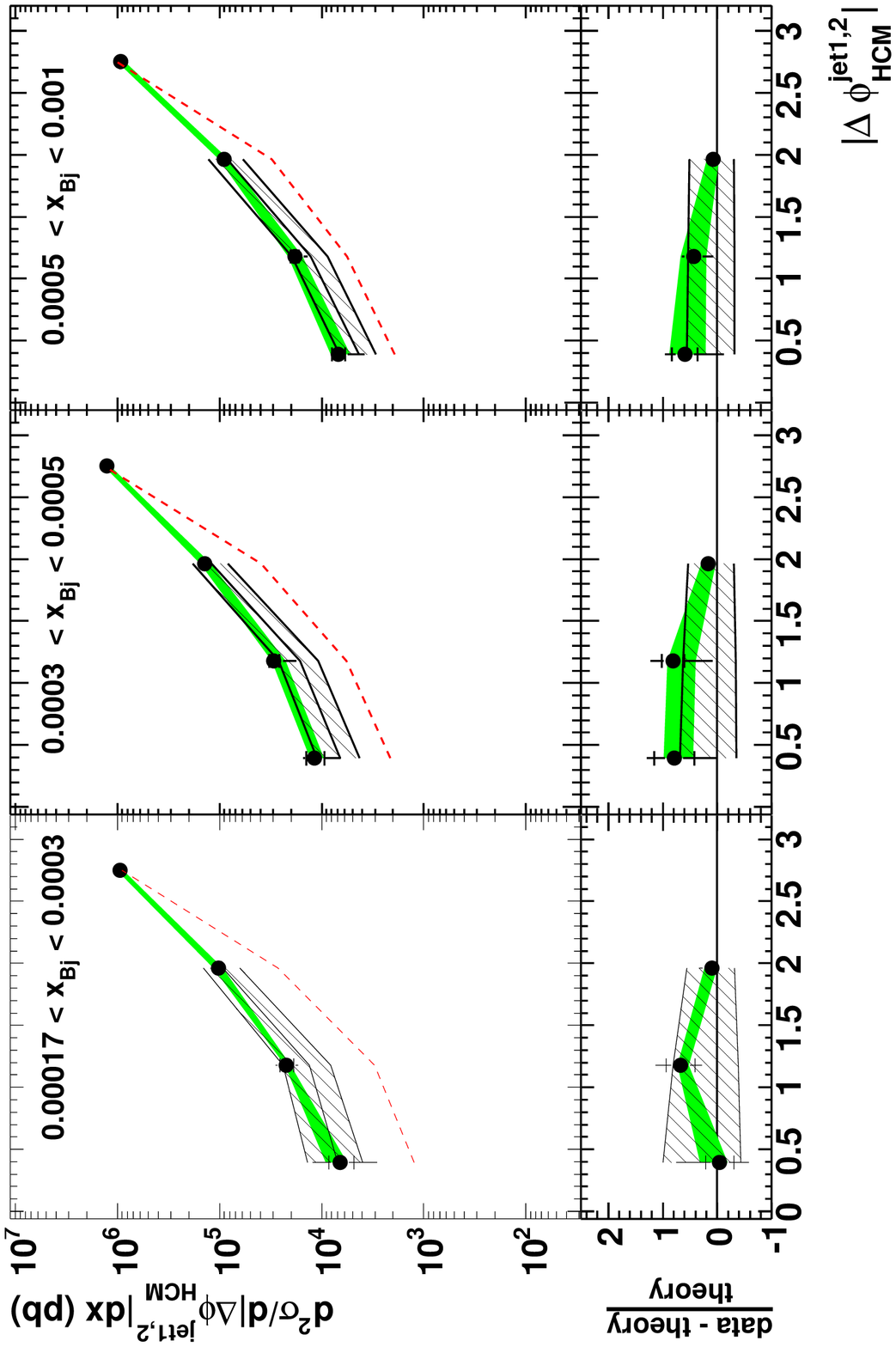}
\caption{(left) Bjorken-x  dependence and (right) azimuth dependence 
of  di-jet distributions at HERA as measured 
by ZEUS\protect\cite{zeus1931}.} 
\label{fig:phizeus}
\end{figure}

DIS multi-jet distributions 
associated with $Q^2 > 10$ GeV$^2$ 
and $10^{-4} < x < 10^{-2}$ have recently 
been measured by 
the {\small ZEUS} collaboration~\cite{zeus1931}, 
and compared with 
next-to-leading-order   calculations~\cite{nagy}. 
Results  for di-jet distributions are 
shown in   Fig.~\ref{fig:phizeus}~\cite{zeus1931}. 
The plot on the left in Fig.~\ref{fig:phizeus} shows the 
$x$-dependence of the  di-jet distribution integrated  over 
$\Delta \phi < 2 \pi / 3$, where $\Delta \phi$ is the 
azimuthal separation 
between the two high-$E_T$ 
jets. The  plot on the right 
shows the di-jet
distribution in $\Delta \phi$  for different bins of $x$.
The overall agreement of data  with NLO results is within 
errors. 
However, the variation of the 
predictions from  order-$\alpha_s^2$ to order-$\alpha_s^3$ 
is  significant. Given the large difference between 
order-$\alpha_s^2$ and order-$\alpha_s^3$ results, 
the theoretical 
uncertainty at NLO  appears to be 
 underestimated 
by the error band 
obtained from the conventional method of 
varying the renormalization/factorization 
scale.\footnote{Besides angular correlations,  sizeable 
 uncertainties at NLO also affect other associated distributions 
such as  momentum  correlations.    On the other 
hand,   NLO results 
are  much more stable for  inclusive 
jet cross sections~\cite{zeus1931}.}

The stability of predictions for 
the jet observables in Fig.~\ref{fig:phizeus} 
depends on a number of   physical processes.  Part of these  
concern  nonperturbative dynamics, including 
jet clustering and hadronization\footnote{See~\cite{hj_ang} 
and references therein for more discussion of these effects.},    
part concern radiative corrections  
at higher order. Fixed-order 
calculations beyond NLO are not within present reach 
 for multi-jet processes in hadronic 
collisions.   
 Resummed calculations of 
 contributions from    multiple infrared 
emissions are performed in~\cite{delenda}.  
These contributions are 
  enhanced in  the region where the two high-$E_T$ 
jets are nearly back-to-back, and are also 
taken into account 
by parton-shower methods 
as in  \herwig~\cite{herwref}. 
Note however that sizeable corrections 
in Fig.~\ref{fig:phizeus}  arise  for 
decreasing $\Delta \phi$, where   
the two jets are not close to back-to-back  and 
one has effectively three  
well-separated hard jets~\cite{hj_rec}. 
The corrections increase as $x$ decreases. 
By analyzing   the angular distribution of 
the third jet,  Ref.~\cite{hj_ang} finds  
significant contributions  
 from configurations where 
the transverse momenta in the initial-state 
shower are not ordered. 
These contributions are not included in standard 
parton showers, e.g.  \herwig~\cite{herwref} and  
\pythia~\cite{pythref}. 
The next section describes results of 
parton-shower calculations that take these contributions 
into account.  

\section{Parton showers 
with k$_\perp$-branching and jet correlations}
\label{sec2}

  Corrections to the initial-state shower due to 
non-ordering in k$_\perp$  can be incorporated  
 in  Monte-Carlo  event 
generators~\cite{krauss-bfkl,golec-mc,lonn,junghgs},    
with logarithmic accuracy for $ x \ll 1$,  
by implementing   
  unintegrated parton distributions defined through 
  high-energy factorization~\cite{hef}. 
  Although none of the above generators are 
as developed as standard
 Monte-Carlos  like 
\herwig~\cite{herwref} or   
\pythia~\cite{pythref}, they have the potential
advantage of a more accurate  treatment  of the 
 initial-state radiative  effects that  
the observations of Sec.~\ref{sec1} suggest to be relevant 
 for  multi-jet correlations. 
  
\begin{figure}[htb]
\vspace{55mm}
\includegraphics{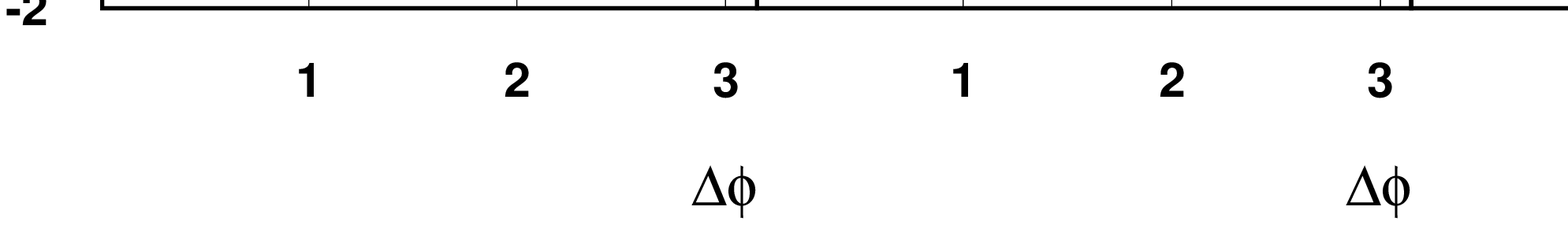}
\includegraphics{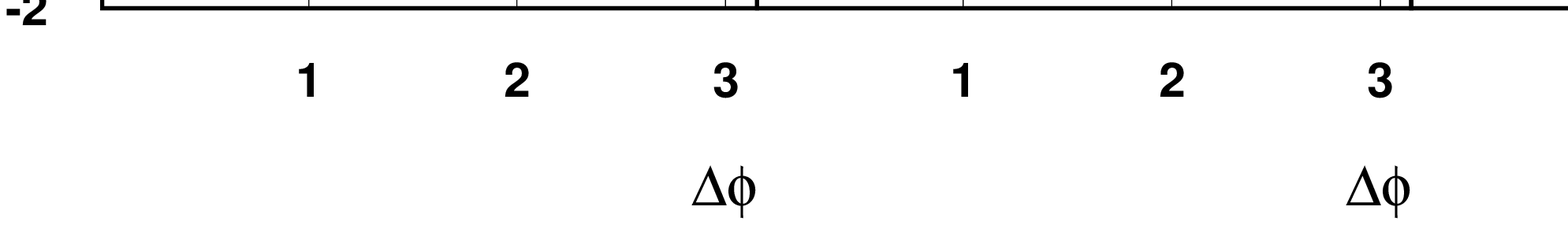}
\caption{Angular jet correlations~\protect\cite{hj_ang}  
obtained by \cascade\ and \herwig, 
compared with 
ZEUS data~\protect\cite{zeus1931}: 
(left) di-jet cross section; (right) three-jet cross section.
} 
\label{fig:phipage}
\end{figure}

Fig.~\ref{fig:phipage} shows  results~\cite{hj_ang}  for
the azimuthal distribution of di-jet and three-jet
cross sections obtained by the 
k$_\perp$-shower Monte-Carlo  \cascade~\cite{junghgs} and by 
\herwig~\cite{herwref},  compared with the
measurement~\cite{zeus1931}.  The shape of the 
 distribution  is different for the two Monte-Carlos. 
  \cascade\  gives large differences from \herwig\
in the region where the azimuthal separations
 $\Delta \phi$ between the
leading jets are small. It becomes closer to \herwig\
as  $\Delta \phi$ increases. This is consistent 
 with the expectation
that both Monte-Carlos give  reasonable approximations
 near the  back-to-back region. 
The description of the
measurement by \cascade\ is
 good, whereas   \herwig\  is not
sufficient to describe the measurement 
in the small $\Delta \phi$ region.

\begin{figure}[htb]
\vspace{43mm}
\includegraphics{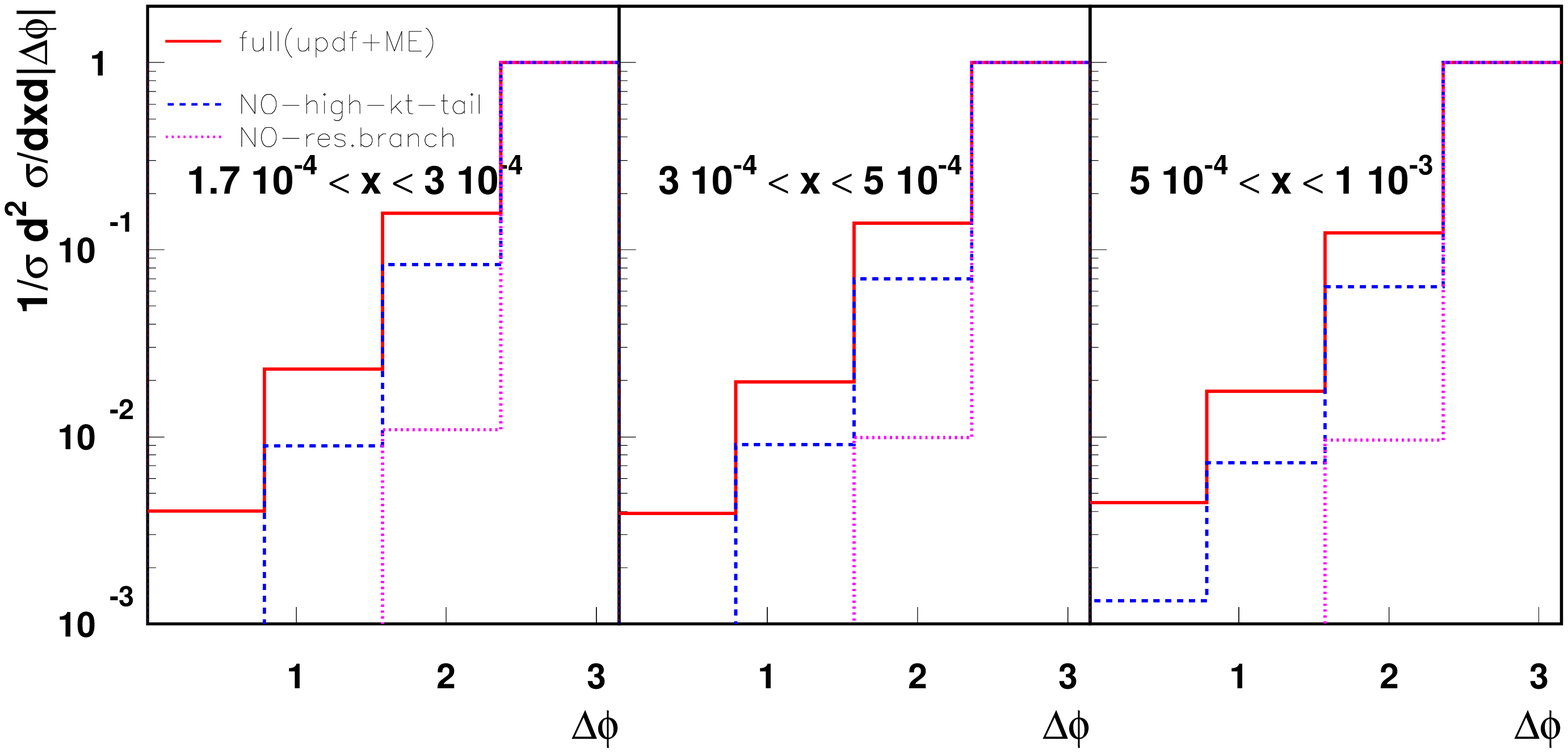}
\caption{Azimuthal distribution normalized to the 
back-to-back cross section~\protect\cite{hj_ang}: 
(solid red)  full result 
(u-pdf $\oplus$ ME); (dashed blue) no finite-k$_\perp$ 
correction in ME 
 (u-pdf $\oplus$ ME$_{collin.}$); 
(dotted violet) u-pdf with no resolved branching.
} 
\label{fig:ktord}
\end{figure}

In the k$_\perp$-shower calculation 
 both the parton distributions (unintegrated pdf's,  
 fitted 
 from experiment) and the hard matrix elements 
 (ME's, computed perturbatively)   are 
 transverse-momentum dependent. 
The parton branching includes 
regions that are not ordered in  k$_\perp$ along the 
initial-state decay chain. 
 Fig.~\ref{fig:ktord} 
shows different approximations to the  
azimuthal dijet distribution normalized to the 
back-to-back cross section. The solid 
red curve is   the full  result.  
  The 
 dashed blue curve is  obtained 
from the same unintegrated pdf's but 
not including the  transverse-momentum dependence of 
the hard ME.  
We see that  the 
high-k$_\perp$ component  in the hard ME 
is essential to describe 
jet correlations particularly 
for small $\Delta \phi$.  The dotted violet curve is    
  obtained from the  
unintegrated pdf 
without any resolved branching, corresponding  
 to nonperturbative, predominantly 
 low-k$_\perp$ modes. The results of 
Fig.~\ref{fig:ktord}  illustrate  that the  
 k$_\perp$-dependence  
in the unintegrated pdf alone 
is not sufficient  to describe jet production 
quantitatively. 

Jet multiplicities and jet momentum  correlations  are analyzed  in~\cite{hj_ang} 
along    lines  similar to those discussed  above for azimuthal distributions. 
The largest differences between  
 \herwig\ and the k$_\perp$-showering are found in the 
contribution of high multiplicities and   correlations  
in the transverse-momentum imbalance between the leading jets.

In summary, the 
 results give  a consistent   parton-shower picture 
of distributions associated to   multi-jet production,   including 
correlations. This  is expressed   in terms of 
  unintegrated pdf's  
 convoluted 
with transverse-momentum dependent   
hard kernels.\footnote{The u-pdf's 
are defined gauge-invariantly~\cite{hef} 
for small $x$. General operator 
definitions,  including the region  $x \sim 1$,  and issues of 
lightcone singularities  are discussed  e.g. 
in~\cite{uop,endp} and references therein.}  
The physical    picture reflects the 
growth of the k$_\perp$ 
transmitted along the initial-state jet,  and includes 
 corrections to the collinear  ordering 
implemented in    standard showers such as 
\herwig\ and \pythia. 
We have  considered  production of jets  
 in the non-forward region. 
 Results for jet correlations are less  dependent on 
details of u-pdf evolution models 
than in the case of forward-region observables. 
Owing to the 
large phase space available for jet production,  
the DIS kinematic region under consideration   
may  be relevant  for 
understanding the extrapolation of initial-state
showering  effects to the LHC, despite the 
lower energy. 

\vskip 0.4 cm 

\noindent  {\bf Acknowledgments.} We thank the organizers of the 
Symposium for the invitation, and gratefully acknowledge 
the hospitality and support of the Galileo Galilei Institute for Theoretical 
Physics.

\end{document}